\documentclass[prd,12pt,nofootinbib,superscriptaddress,reprint]{revtex4-2}
\usepackage{bm}
\usepackage{amsmath}
\usepackage{amssymb}
\usepackage{booktabs}
\usepackage{graphicx}
\usepackage{float}
\usepackage{longtable}
\usepackage{subfigure}
\usepackage{hyperref}
\usepackage{comment}
\usepackage{mathrsfs}
\usepackage{xcolor}
\usepackage{color}
\usepackage{CJKutf8}
\usepackage{ulem}
\usepackage{CJK}
\hypersetup{
	colorlinks=true,
	linkcolor=red,
	citecolor=blue
}

\begin{document}
	
\begin{CJK*}{UTF8}{gbsn}
\title{Primordial black holes and scalar-induced gravitational waves from the polynomial attractor model}
\author{Zhongkai Wang (王中凯)}
\email{ztlu@hust.edu.cn}
\affiliation{School of Physics, Huazhong University of Science and Technology, Wuhan, Hubei
430074, China}
\author{Shengqing Gao (高盛晴)}
\email{gaoshengqing@hust.edu.cn}
\affiliation{School of Physics, Huazhong University of Science and Technology,
Wuhan, Hubei
430074, China}
\author{Yungui Gong (龚云贵)}
\email{Corresponding author. yggong@hust.edu.cn}
\affiliation{School of Physics, Huazhong University of Science and Technology, Wuhan, Hubei
430074, China}
\affiliation{Department of Physics, School of Physical Science and Technology, Ningbo University, Ningbo, Zhejiang 315211, China}
\author{Yue Wang (王岳)}
\email{ywang123@hust.edu.cn}
\affiliation{School of Physics, Huazhong University of Science and Technology, Wuhan, Hubei
430074, China}

\begin{abstract}
Primordial black holes (PBHs) generated in the early Universe are considered as one of the candidates for dark matter. 
To produce PBHs with sufficient abundance, the primordial scalar power spectrum needs to be enhanced to the order of 0.01. 
Considering the third-order polynomial potential with polynomial attractors, 
we show that PBHs with a mass of about $10^{17}$g can be produced while satisfying the constraints from the cosmic microwave background observations at the 2$\sigma$ confidence level. 
The mass of PBHs produced in the polynomial  attractors can be much bigger than that in the exponential $\alpha$ attractors.
By adding a negative power-law term to the polynomials, 
abundant PBHs with different masses and the accompanying scalar-induced gravitational waves (SIGWs) with different peak frequencies are easily generated. 
The PBHs with masses around $10^{-15}\textbf{--} 10^{-12}$ $M_\odot$ can account for almost all dark matter.
The SIGWs generated in the nanohertz band can explain the recent detection of stochastic gravitational-wave background by the pulsar timing array observations. 
The non-Gaussianities of the primordial curvature perturbations in the squeezed and equilateral limits are calculated numerically. 
We find that the non-Gaussianity correction enhances the PBH abundance
which makes the production of PBHs much easier,
but the effect of non-Gaussianity on the generation of SIGWs is negligible.
\end{abstract}

\maketitle

\end{CJK*}

\section{Introduction}
The direct detection of gravitational waves (GWs) by the Laser Interferometer Gravitational-Wave Observatory(LIGO) and Virgo Collaborations opened a new window to test gravity and understand the Universe \cite{LIGOScientific:2016aoc,LIGOScientific:2016emj,LIGOScientific:2018mvr,LIGOScientific:2020ibl,LIGOScientific:2021usb,LIGOScientific:2021djp}.
In particular, some of the binary black holes observed may consist of primordial black holes (PBHs) \cite{Bird:2016dcv,Sasaki:2016jop,Inomata:2016rbd,DeLuca:2020sae,DeLuca:2021wjr,Franciolini:2021tla},
 and the detections of stochastic GW background by pulsar timing arrays (PTAs)
also hint at the existence of PBHs \cite{DeLuca:2020agl, Vaskonen:2020lbd, Kohri:2020qqd, Domenech:2020ers, Atal:2020yic,Yi:2021lxc,NANOGrav:2023hvm,EPTA:2023xxk,Yi:2023mbm,Franciolini:2023pbf,Ellis:2023oxs,Basilakos:2023xof,Basilakos:2023jvp}.
Furthermore, PBHs can be regarded as one of the candidates for dark matter (DM) and their abundance and mass ranges are tightly constrained by  observations 
\cite{Ivanov:1994pa,Khlopov:2004sc,Frampton:2010sw,Belotsky:2014kca,Carr:2016drx,Carr:2020xqk}.

If overdense inhomogeneities seeded from the primordial curvature perturbations at the horizon reentry exceed the threshold value $\delta_c$ during radiation domination,
then  PBHs could form in the overdense regions through gravitational collapse \cite{Carr:1975qj, Hawking:1971ei, Carr:1974nx}.
Accompanied by the generation of PBHs,
the large curvature perturbation also produces scalar-induced GWs (SIGWs) that contribute to the stochastic gravitational-wave background (SGWB) \cite{Ananda:2006af,Baumann:2007zm, Saito:2008jc,Kohri:2018awv,Espinosa:2018eve, Cai:2019elf, Inomata:2019ivs, Domenech:2020kqm, Braglia:2020taf, Yi:2022anu,Domenech:2021ztg}.
To generate a significant abundance of PBHs, the amplitude of the power spectrum $\mathcal{P}_\zeta$ of the primordial curvature perturbation $\zeta$ should be $A_s\sim\mathcal{O}(0.01)$ at small scales 
\cite{Ballesteros:2017fsr, Dalianis:2019vit, Sato-Polito:2019hws,Lu:2019sti, Ballesteros:2020qam,  Fu:2019ttf, Lin:2020goi, Yi:2020kmq, Yi:2020cut, Ragavendra:2020sop, Kawaguchi:2022nku, Aldabergenov:2023yrk, Cheung:2023ihl, Hooper:2023nnl,Papanikolaou:2022did,Kubota:2023ked, Di:2017ndc, Garcia-Bellido:2017mdw, Ezquiaga:2017fvi, Germani:2017bcs, Espinosa:2017sgp,Ballesteros:2018wlw,Gao:2018pvq,Sasaki:2018dmp,Passaglia:2018ixg}.
Compared with the constraint on the amplitude of the power spectrum at the cosmic microwave background (CMB) scales $A_s \approx 2.1\times10^{-9}$ \cite{Planck:2018jri},
the power spectrum at small scales has to be enhanced by at least 7 orders of magnitude.

To amplify the power spectrum at small scales, 
a straightforward way is to construct a flat plateau, i.e.,  
an inflection point in the canonical inflation potential 
\cite{Garcia-Bellido:2017mdw,Ezquiaga:2017fvi,Germani:2017bcs, Espinosa:2017sgp, Ballesteros:2018wlw, Gao:2018pvq,Sasaki:2018dmp,Passaglia:2018ixg,Di:2017ndc, Martin:2012pe,Motohashi:2014ppa,Motohashi:2017kbs,Bezrukov:2017dyv,Cicoli:2018asa,Passaglia:2019ueo,Bhaumik:2019tvl,Xu:2019bdp,Dalianis:2018frf},
leading to the so-called ultraslow-roll (USR) \cite{Tsamis:2003px,Kinney:2005vj} stage during inflation. 
From the fact that an $m$-order polynomial has $m-2$ inflection points, 
the potential with $m=3$ can have one inflection point in the inflaton potential. 
Thus, there are some attempts to combine the exponential $\alpha$ attractor models \cite{Starobinsky:1980te, Salopek:1988qh, Kallosh:2013hoa, Kallosh:2013lkr, Kallosh:2013maa,Kallosh:2013daa, Kallosh:2013tua, Kallosh:2013yoa, Galante:2014ifa,Kallosh:2022feu} with the third-order polynomial $V(\psi)=V_0\left[1+c_1f(\psi)+c_2f^2(\psi)+c_3f^3(\psi)\right]^2$ to amplify the power spectrum and generate abundant PBHs \cite{Dalianis:2018frf, Iacconi:2021ltm, Frolovsky:2022qpg, Frolovsky:2023hqd}. 
However, the masses of PBHs produced from these models are too small. 
Adding a negative power-law term $f^{-2}(\psi)$ can solve this problem \cite{Frolovsky:2023hqd}. 

Apart from the exponential $\alpha$ attractors, there are also polynomial $\alpha$ attractors \cite{Dvali:1998pa, Burgess:2001fx, Lorenz:2007ze, Kallosh:2018zsi,Kallosh:2022feu}.
Extending the polynomial $\alpha$ attractors to a hybrid polynomial attractor model, 
the possibility of PBH production and SGWB generation was studied with a two-stage inflation \cite{Braglia:2022phb}.
In this paper, we replace the exponential $\alpha$ attractors with the polynomial $\alpha$ attractors in the third-order polynomial potential with and without the additional negative power-law term. 

Near the inflection point, the slow-roll (SR) conditions are violated \cite{Motohashi:2017kbs}
and there may exist a non-negligible non-Gaussianity.
The abundance of PBHs and the energy density of accompany SIGWs
can be affected by large non-Gaussianity of the primordial curvature perturbations \cite{Saito:2008em,Franciolini:2018vbk,Cai:2018dig,Kehagias:2019eil,Atal:2018neu,Riccardi:2021rlf,Zhang:2020uek, Zhang:2021vak,Gao:2020tsa,Gao:2021vxb}.
The impacts of non-Gaussianity of the primordial curvature perturbation $\zeta$ on the formation of PBHs were discussed in the literature \cite{Bullock:1996at, Ivanov:1997ia, Yokoyama:1998xd,Ferrante:2022mui,Atal:2021jyo,Choudhury:2023kdb,Choudhury:2023fwk}.  
There are some methods to calculate the PBH fractional energy density $\beta$ in the presence of non-Gaussianity, such as changing the variables in the Gaussian probability density function \cite{Byrnes:2012yx} and adopting the path-integral formulation \cite{Franciolini:2018vbk,Saito:2008em}. 
In this paper, we use the latter method to calculate $\beta$.

This paper is organized as follows. In Sec. II, we show the models in detail and present the power spectra of the models. In Sec. III, we discuss the PBH abundance produced from the models. In Sec. IV, we compute the generation of SIGWs during radiation domination.
We also compare the SIGWs produced in the models with the recent detection of nanohertz stochastic GW background by the North American Nanohertz Observatory for Gravitational Waves (NANOGrav) \cite{NANOGrav:2023ctt} and the European Pulsar Timing Array (EPTA) \cite{EPTA:2023fyk}. 
In Sec. V, we calculate the non-Gaussianities of one model in the squeezed and equilateral limits
and discuss the effects of non-Gaussianity on the production of PBHs and SIGWs.
We conclude the paper in Sec. VI.

\section{The Models}
For the enhancement of the primordial scalar power spectrum at small scales with $\alpha$ attractors,
third-order polynomial potentials with an inflection point
    \begin{equation}
    \label{potentiale}    V(\psi)=V_0\left[1+c_1f(\psi)+c_2f^2(\psi)+c_3f^3(\psi)\right]^2
    \end{equation}
were usually discussed \cite{Dalianis:2018frf, Iacconi:2021ltm, Frolovsky:2022qpg, Frolovsky:2023hqd}.
For T models, $f=f_t(\psi)=\tanh{(\psi/\sqrt{6\alpha})}$ \cite{Dalianis:2018frf, Iacconi:2021ltm};
for E models, $f=f_e(\psi)=\exp{(-\sqrt{2/3\alpha}\,\psi)}$ \cite{Frolovsky:2022qpg},
this also belongs to exponential $\alpha$ attractors \cite{Kallosh:2022feu}.
The masses of PBHs produced by T models with the potential \eqref{potentiale} are smaller than $10^8$ g \cite{Iacconi:2021ltm}, and the PBHs produced by E models with the potential \eqref{potentiale} are asteroid-sized black holes \cite{Frolovsky:2022qpg}.
To obtain PBHs with larger masses, one must decrease the spectral tilt $n_s$ \cite{Frolovsky:2023hqd}.
Because of the observational constraint by Planck observations,
$n_s=0.9649\pm 0.0042$ (68\% confidence level) \cite{Planck:2018jri},
the spectral tilt $n_s$ cannot decrease too much.
To overcome this problem, a negative power-law term $f^{-2}(\psi)$ was added to the potential for E models with extreme fine-tuning of the parameters \cite{Frolovsky:2023hqd}.

On the other hand, the polynomial potential with inverse powers 
\begin{equation}
\label{invpow1}    
V(\psi)=V_0\left(1-\frac{\mu^k}{\psi^k}+\cdots \right)
\end{equation}
gives the polynomial $\alpha$ attractors \cite{Kallosh:2022feu}
\begin{equation}
\label{nsr1} 
n_s=1-\frac{2}{N}\frac{k+1}{k+2}, \quad 
r=\frac{(4k)^{\frac{2}{k+2}}(3\alpha)^{\frac{k}{k+2}}}{[(k/2+1)\mathit{N}]^{2-\frac{2}{k+2}}},
\end{equation}
where $N$ is the total number of e-folds before the end of inflation when the pivotal scale $k_*$ exits the horizon, and 
$r$ is the tensor-to-scalar ratio, $\alpha=2\mu^2/3$.

In this paper, we use the polynomial attractor \cite{Kallosh:2022feu} to discuss the enhancement of the primordial scalar power spectrum at small scales from the polynomial potential \eqref{potentiale} with the addition of one negative power-law term,
\begin{equation}
\label{polypot1}
\begin{split}
V(\psi)=V_0[1+c_4f_n^{-1}(\psi)+c_1f_n(\psi)\\
+c_2f_n^2(\psi)
+c_3f_n^3(\psi)]^2,
\end{split}
\end{equation}
where the function $f_n(\psi)=\psi^{-n}$ and we choose $n=3$ for the discussion of PBH production and SGWB generation.
The factor $(k+1)/(k+2)$ in Eq. \eqref{nsr1} helps to increase the value of $n_s$,
so the usual SR stage can be shorter and it is possible to get larger mass PBHs with the polynomial $\alpha$ attractors than with T and E models.
To understand why the model works and how to choose the model parameters,
we divide the total number of e-folds $N$ into two parts, 
the first part is the SR stage with the  number of e-folds $N_\text{sr}$,
the second part is the USR stage until the end of inflation with the number of e-folds $N-N_\text{sr}$.
Since the polynomial attractors \eqref{nsr1} are obtained for SR inflation, 
in the discussion of observational constraints by Planck, 
we should use $N_\text{sr}$ in Eq. \eqref{nsr1}; 
then we use the result to guide us to choose model parameters.
Note that the division into SR and USR stages is not exact and is just a rough approximation, so
the attractors \eqref{nsr1} determined by the SR stage are not the exact predictions of the model,
and the results for $n_s$ and $r$ are actually obtained by numerical calculation.

For the potential (\ref{potentiale}), the inflection point (IP) is
\begin{equation}
\label{ip1}
    f(\psi^\text{IP})=-\frac{c_2}{3c_3},
\end{equation}
and the derivative of the potential $V(f(\psi))$ at the inflection point is
\begin{equation}
\label{inflpt1}
    V'(f(\psi^\text{IP}))=c_1-\frac{c_2^2}{3c_3}.
\end{equation}
If we choose the value of $c_1$ so that the derivative of the potential at the inflection point $\psi^\text{IP}$ is exactly zero, 
then usually we will get a very large number of e-folds. 
To avoid a big number of e-folds $N$, 
we add a small number $\delta$ to $c_1$ so that the derivative of the potential at the inflection point $\psi^\text{IP}$ is close to zero, but not exactly zero, 
i.e., we introduce $\delta$ to adjust the value of $N$,
\begin{equation}
\label{polypot3}
c_1=\frac{c_2^2}{3c_3}+\delta, \qquad \delta\ll 1.
\end{equation} 

To compare the enhancement of the power spectrum between exponential $\alpha$ attractors and polynomial $\alpha$ attractors, 
we first discuss the difference between the results for the potential \eqref{potentiale} with $f=f_e$ and $f=f_n$,
then we will discuss the power spectrum of the polynomial $\alpha$ attractor with the potential \eqref{polypot1}.

\subsection{Exponential and polynomial attractors}

There are three parameters in the polynomial potential \eqref{potentiale}.
 The condition \eqref{inflpt1} for the inflection point reduces one parameter, so there are two free parameters left.
 At large $\psi$, the potential (\ref{potentiale}) is dominated by the linear term  $c_1 f(\psi)$ and the results depend on the parameter $c_1$ only because of the existence of attractors.
 The value of $V_0$ is determined by the amplitude of the scalar power spectrum.

To show how to determine the model parameters, we take the model $f(\psi)=f_2=\psi^{-2}$ as an example.
Even though the attractors \eqref{nsr1} are independent of $c_1$,
the value of the scalar field $\psi_*$ at the pivot scale $k_*=0.05$ Mpc$^{-1}$ is $\psi_*\approx (-16 c_1 N_\text{sr})^{1/4}$, 
so $\psi_*$ is determined by $c_1$ and the e-folds $N_\text{sr}$ for the SR stage.
Because of the attractors \eqref{nsr1} and the CMB constraints on $n_s$ and $r$, 
$N_\text{sr}$ cannot be too small. 
In order to enhance the power spectrum at small scales, 
$N_\text{sr}$ cannot be too large either.
For the convenience of discussion on the choices of parameters and the rough estimation of the observables at large scales, 
we take a fiducial $N_\text{sr}=40$. 
Note that the actual value of $N_\text{sr}$ is obtained by numerical calculation and is different for different models and model parameters.
For fixed $N_\text{sr}$, $\psi_*$ is determined by $c_1$.
The value of $\psi_*$ is larger if $|c_1|$ is larger.
Once we take a value of $c_1$, then $c_2$ and $\psi^\text{IP}$ can be determined from $c_1$ and $c_3$.
Combining Eqs. \eqref{ip1} and \eqref{polypot3}, we get $\psi^\text{IP}=(3c_3/c_1)^{1/4}=2(-3c_3N_\text{sr})^{1/4}/\psi_*$,
so $\psi^\text{IP}$ is smaller if $|c_1|$ is larger.
Now we fix $c_3=-1.2$ and $N_\text{sr}=40$ to get $\psi_*$ by 
choosing $c_1=-1.7$, $-1.9$, $-2.1$, respectively, 
then we adjust the value of $\delta$ 
and numerically solve the background and perturbed cosmological equations
to obtain the scalar power spectrum $\mathcal{P}_\zeta$ of the primordial curvature perturbation $\zeta$, 
the scalar spectral tilt $n_s$, and the tensor-to-scalar ratio $r$ at the pivot scale $k_*$. 
The results show that, as $|c_1|$ becomes larger, 
$\psi_*$ and $r$ are bigger, but $N$ and $k^\text{peak}$ become smaller.
For bigger $|c_1|$, the power spectrum can be enhanced to the order of $0.01$ with a smaller number of e-folds,
so the scale at the end of inflation is also smaller, 
which leads to a smaller peak scale $k^\text{peak}$ even though $\psi^\text{IP}$ is smaller.
Therefore, to get massive PBHs, we should take the bigger value of $|c_1|$.
However, if $|c_1|$ is too large, then inflation ends before the inflection point and the enhancement cannot happen. 
On the other hand, if $|c_1|$ is too small, then the inflaton stays at the inflection point for a much longer time, leading to a large number of $N$.
Since $N$ becomes smaller for larger $|c_1|$, we can increase $N_\text{sr}$ to get bigger $n_s$.
Even though we get bigger $n_s$ and smaller $r$, 
the peak scale $k^\text{peak}$ becomes larger and the mass of generated PBHs becomes smaller,
so we still use smaller $N_\text{sr}$.

Now we fix $c_1=-2.1$ and $\psi_*=6.055$, then vary the value of $c_3$. 
The results tell us that $N$, $n_s$, and $k^\text{peak}$ are smaller with smaller $|c_3|$.
Again, if we increase $\psi_*$ to get larger $n_s$ for the same $c_3$, $k^\text{peak}$ becomes bigger.
The above discussion tells us that we need to choose larger $|c_1|$ and smaller $|c_3|$ to get smaller $k^\text{peak}$. 

To make a comparison for the models $f_e$ and $f_n$ with different $n$,
we fix $n_s=0.9565$, which is consistent with the Planck constraint at the $2\sigma$ confidence level.
Following the procedure discussed above, 
we can find the appropriate parameters for each model, then we calculate the power spectra, and the results are shown in Table \ref{tab:f1fe}.
Even though we take the lower limit of the $2\sigma$ confidence level of the Planck results, 
$k^\text{peak}$ can only be as small as $10^{14}\sim 10^{15} \text{ Mpc}^{-1}$ for the models $f_n$,
and $k^\text{peak}$ is about $10^{18} \text{ Mpc}^{-1}$ for the model $f_e$.
Furthermore, for the polynomial attractors $f_n$, $N$ is smaller than $40$. 
If we increase $\psi_*$, $N$ and $n_\text{s}$ can be bigger, but then $k^\text{peak}$ becomes bigger too.
On the other hand, the results in Table \ref{tab:f1fe} show that $k^\text{peak}$ in the model $f_n$ can be smaller than that in the model $f_e$, 
so we only consider the polynomial attractor $f_3$ in the following discussions.

\begin{table*}
\centering
\begin{tabular}{cccccccccc} \toprule
Models  & $V_0$ &$c_1$ & $c_3$  & $\delta$      &  $\psi_*$  & $N$ & $r$ & $\mathcal{P}_\zeta^\text{peak}$  &  $k^\text{peak}$\\ \midrule
$f_1$   & 3.3$\times10^{-9}$ & -2.9 & -1.16  &  $2.84193\times10^{-2}$    & 8.787    & 39.4 & 0.0503 & 0.011 & 2.8$\times10^{14}$\\
$f_2$   & 5.5$\times10^{-10}$  &  -2.5  & -1.16  &      $8.48823\times10^{-2} $  & 6.113  
& 37.8 & 0.0147 & 0.01 & 4.3$\times10^{14}$\\
$f_3$   & 1.8$\times10^{-10}$  & -2.2 &  -1.17  &  $1.284273\times10^{-1}$  & 4.729   & 38.8 & 0.0053 & 0.014 & 1.6$\times10^{15}$\\
$f_e$   & 1.25$\times10^{-10}$ & -2.6 & -1.15  & $3.74916\times10^{-2}$   & 5.461   & 51.1 & 0.0037 & 0.010 & 7.2$\times10^{18}$\\
				\bottomrule
\end{tabular}
\caption{The parameters and results of different models. The unit of $k^\text{peak}$ is Mpc$^{-1}$.}
\label{tab:f1fe}
\end{table*}

\begin{table*}
\centering
\begin{tabular}{cccccccccccc} 
\toprule
Models & $V_0 $ & $\delta$ &$c_1$ &$c_4$ &  $\psi_*$  &$N$&$n_s$ & $\alpha_s$ &$r$ & $\mathcal{P}_\zeta^\text{peak}$ & $k^\text{peak}$ \\ \midrule
A1 & 8.33$\times10^{-10}$ & 1.1007$\times10^{-3}$  & -1.07 &3.2$\times10^{-4}$ & 3.75   &63.2 &0.9658& -0.0041 &0.025 & 0.0338 & 3.0$\times10^{8}$\\
A2  &  3.53$\times10^{-10}$ & 6.3843$\times10^{-4}$   &  -1.1 & 1.3$\times10^{-4}$ & 4.0   & 54.8 & 0.9652& -0.0015 & 0.011 & 0.029 & 1.7$\times10^{12}$\\
A3 & 2.75$\times10^{-10}$ & 7.574$\times10^{-4}$  & -1.12 & 9.5$\times10^{-5}$& 4.08  &52.4&0.9655 & -0.0017  &0.008 & 0.0269 & 3.6$\times10^{13}$\\
\bottomrule
\end{tabular}
\caption{The parameters and results of modified polynomial attractors. The unit of $k^\text{peak}$ is Mpc$^{-1}$.}
\label{tab:A123}
\end{table*}

\subsection{Modified polynomial attractors}

As discussed above, it is very difficult to produce bigger PBHs with the third-order polynomial potentials \eqref{potentiale},
so we consider the modified polynomial potential \eqref{polypot1}. 
We choose three sets of parameters for the model $f_3$ and denote them as A1, A2 and A3, respectively.

\begin{figure}[htbp]
\includegraphics[width=0.9\columnwidth]{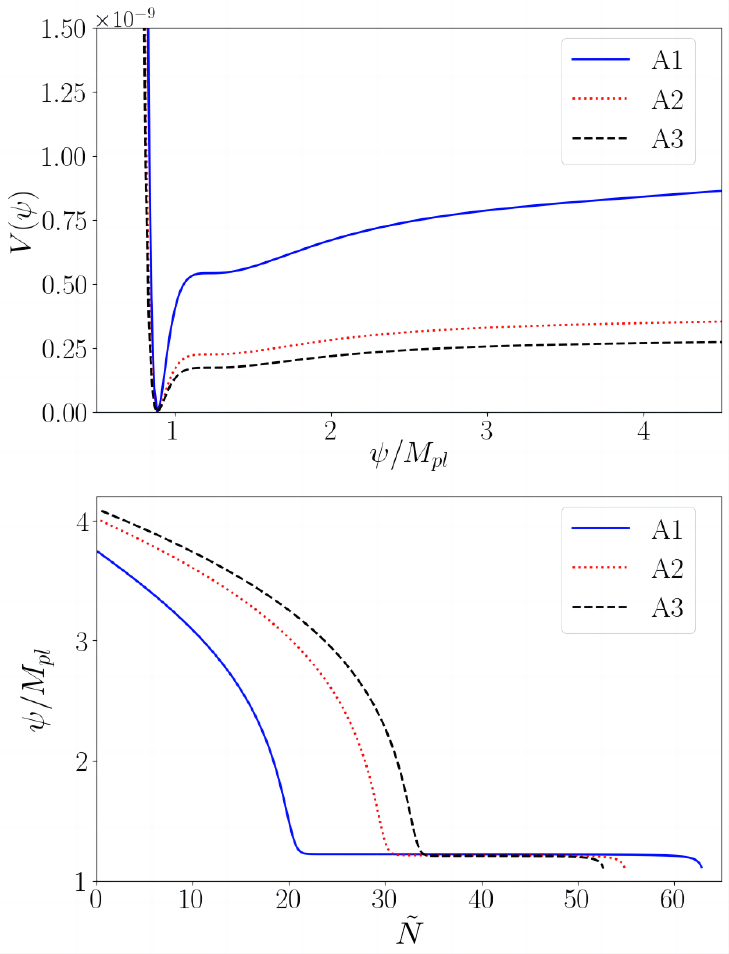}
\caption{Upper: potentials of the models A1, A2, and A3. 
Lower: evolution of the scalar field $\psi$ in terms of the number of e-folds $\tilde{N}$ since the pivotal scale $k_*$ exits the horizon in these models.
Blue solid line, model A1; red dotted line, model A2; black dashed line, model A3.}
\label{vphi_img}

\end{figure}

\begin{figure}
\centering
\includegraphics[width=0.9\linewidth]{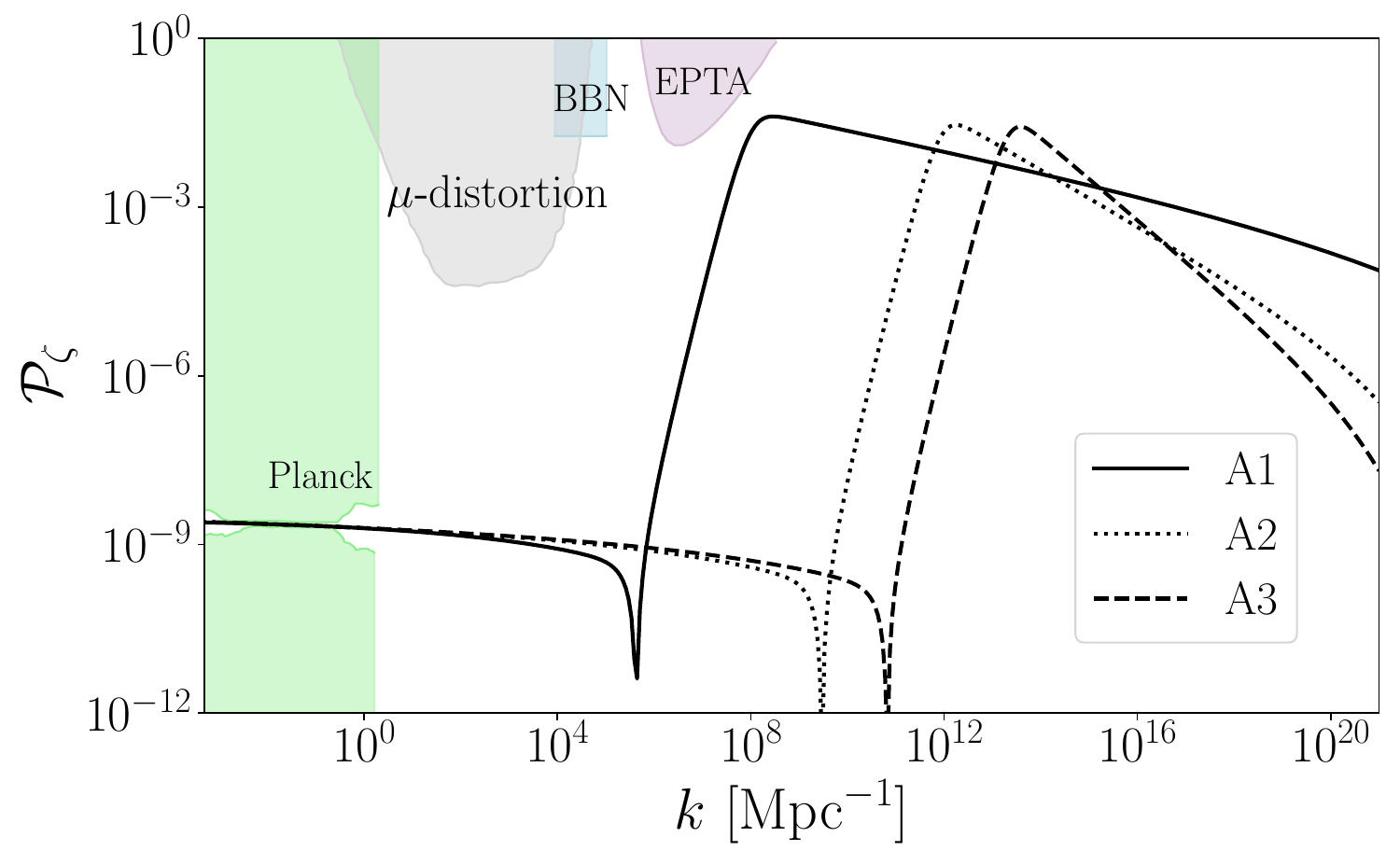}
\caption{The results of the power spectra for the models A1, A2, and A3.
Black solid line, model A1;
black dotted line, model A2;
black dashed line, model A3.
The light green shaded region is excluded by the CMB observations \cite{Planck:2018jri}. 
The light purple, cyan, and gray regions show the constraints from the PTA observations \cite{Inomata:2018epa}, the effect on the ratio between neutron and proton during the big bang nucleosynthesis (BBN) \cite{Inomata:2016uip} and $\mu$-distortion of CMB \cite{Fixsen:1996nj}, respectively.}
\label{f+ps_img}
\end{figure}

We take $c_3=-1.2$ for the models A1, A2, and A3, and the values of the other parameters are shown in Table \ref{tab:A123}.
Note that the $c_4$ term will affect the inflection point (\ref{ip1}) and the relation (\ref{polypot3}), 
but its effect is very small near the inflection point, so we still use Eq. (\ref{polypot3}) to determine $c_2$ and absorb the effect of $c_4$ term into $\delta$.
We plot the potentials of the models and the evolution of the scalar field $\psi$ in terms of the number of e-folds $\tilde{N}$ since the pivotal scale $k_*$ exits the horizon in Fig. \ref{vphi_img}.
From Fig. \ref{vphi_img}, we see that the inflaton stays at the inflection point for about $15\textbf{--}40$ e-folds.
The scalar power spectra are shown in Fig. \ref{f+ps_img}.
From Fig. \ref{f+ps_img}, we see that the power spectrum for model A1 is broad and the power spectra for models A2 and A3 are narrow.
The results for the total number of e-folds $N$, the scalar spectral tilt $n_\text{s}$,
the running of the scalar spectral tilt $\alpha_s$,
the tensor-to-scalar ratio $r$, 
the peak value of the power spectra $\mathcal{P}^\text{peak}_\zeta$, and the peak scale $k^\text{peak}$ are also shown in Table \ref{tab:A123}.
From Table \ref{tab:A123}, we see that 
$n_s$, $r$, and $\alpha_s$ are consistent with the recent CMB constraints \cite{Planck:2018jri}, 
and the peak scales $k^\text{peak}$ are much smaller compared with those in Table \ref{tab:f1fe}.
In the model A1, we even get $k^\text{peak}=3\times 10^8$ Mpc$^{-1}$.
As the peak scale $k^\text{peak}$ becomes smaller, the tensor-to-scalar ratio $r$ becomes larger,
and $r$ reaches $0.025$ in the model A1, so the peak scale $k^\text{peak}$ cannot be very small while satisfying the CMB constraints.
As tighter constraints on $r$ and $\alpha_s$ are expected in the future CMB observations such as CMB-S4 \cite{CMB-S4:2016ple} and LiteBIRD \cite{Matsumura:2013aja}, 
the allowed lower value of the peak scale $k^\text{peak}$ will be limited by future data.

\section{PBH formation}
\begin{figure}
\centering
\includegraphics[width=0.9\linewidth]{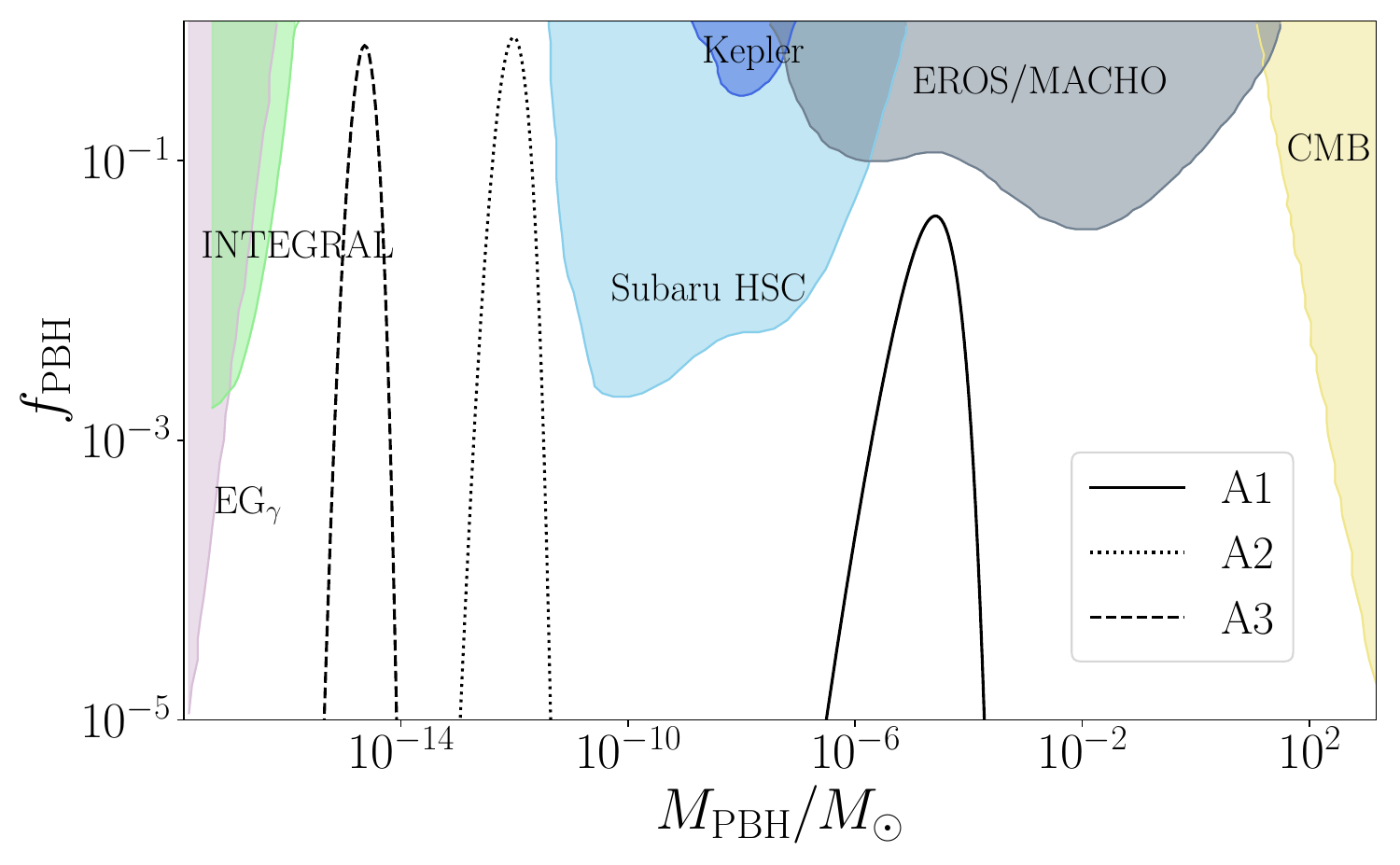}
\caption{The abundance of PBHs produced by the models A1, A2, and A3.
Black solid line, model A1;
black dotted line, model A2;
black dashed line, model A3.
The shaded regions are the observational constraints on PBH abundance: 
the light purple region from extragalactic $\gamma$ rays by PBH evaporation (EG$\gamma$) \cite{Carr:2009jm}, 
the green region from the Galactic Center 511 keV $\gamma$-ray line (INTEGRAL) \cite{Dasgupta:2019cae}, 
the cyan region from microlensing events with Subaru HSC \cite{Niikura:2017zjd}, 
the deep blue region from the Kepler satellite \cite{Griest:2013esa}, 
the gray region from the EROS/MACHO \cite{EROS-2:2006ryy}, 
and the khaki region from accretion constraints by CMB \cite{Poulin:2017bwe}.}
\label{f+pbh_img}
\end{figure}

The overdense region generated by large primordial curvature perturbations at small scales may form PBHs through gravitational collapse after the horizon reentry during radiation domination. 
The mass $M$ of generated PBHs is of the same order as the horizon mass $M_H=(2GH)^{-1}$, $M=\gamma M_H$, where we choose $\gamma=0.2$ \cite{Carr:1975qj}. 
Given a scale $k$, the PBH mass $M$ is \cite{Carr:2016drx,Yi:2020cut}
\begin{equation}
M(k)=3.68\left(\frac{\gamma}{0.2}\right)\left(\frac{g_*}{10.75}\right)^{-1/6}\left(\frac{k}{10^6\text{ Mpc}^{-1}}\right)^{-2}M_{\odot},
\end{equation}
where $M_{\odot}$ is the solar mass, $g_*$ is the effective degrees of freedom at the PBH formation, 
and we do not distinguish the effective degrees of freedom in the entropy and energy density. 
The current fractional energy density of PBHs with mass $M$ to DM is \cite{Di:2017ndc}
\begin{equation}\label{fpbh}
\begin{split}
    f_{\text{pbh}}(M)=&\frac{\beta(M)}{3.94\times10^{-9}}\left(\frac{\gamma}{0.2}\right)^{1/2}\left(\frac{g_*}{10.75}\right)^{-1/4}\\
    &\times\left(\frac{0.12}{\Omega_{\text{DM}}h^2}\right)\left(\frac{M}{M_{\odot}}\right)^{-1/2},
\end{split}
\end{equation}
where $\Omega_{\text{DM}}$ is the current energy density of DM,  
and the Hubble constant $H_0=100h$ km/s/Mpc.
The PBH fractional energy density $\beta(M)$ at the formation for Gaussian curvature perturbation $\zeta$ is \cite{Young:2014ana}
\begin{equation}
\beta^G(M)\approx\sqrt{\frac{2}{\pi}}\frac{\sigma_{R}(M)}{\delta_c}\mathrm{exp}\left(-\frac{\delta_c^2}{2\sigma^2_{R}(M)}\right),
\end{equation}
where $\delta_c$ is the critical density perturbation for the PBH formation.
The mass variance $\sigma_{R}(k)$ on the comoving scale $k=aH=R^{-1}$ is \cite{Young:2014ana}
\begin{equation}\label{sigmaR2}
\sigma^2_{R}(k)=\bigg(\frac{4}{9}\bigg)^2\int\frac{dq}{q}W^2(q/k)(q/k)^4\mathcal{P}_{\zeta}(q),
\end{equation}
and the Gaussian window function $W(x)=\exp(-x^2/2)$. 
In this paper, we choose the parameters as $\delta_c=0.4$ \cite{Musco:2012au,Harada:2013epa,Tada:2019amh,Escriva:2019phb,Escriva:2019phb,Yoo:2020lmg}, $g_*=106.75$, and $\Omega_{\text{DM}}h^2=0.12$. 
Using the power spectra given in Fig. \ref{f+ps_img}, 
we numerically compute the abundance of PBHs with the above formulas and the results for the models A1, A2, and A3 are shown in Fig. \ref{f+pbh_img}.
The results show that all the produced PBHs satisfy the observational constraints.
The peak values of the mass and the abundance of PBH produced at the peak scale $k^\text{peak}$ are shown in Table \ref{tab:pbh}.
As seen from Table \ref{tab:pbh},
in the models A1, A2, and A3, the PBH masses are around $10^{-5}M_\odot$, $10^{-12}M_\odot$, and $10^{-15}M_\odot$, respectively.
PBHs with masses around $10^{-12}M_\odot$ and $10^{-15}M_\odot$ can account for most of DM.
\begin{table}
\centering
\begin{tabular}{cccc} 
\toprule
Models & $M^\text{peak}/M_\odot$ & $f_\text{pbh}$ &$f_\text{c}$/Hz  \\ \midrule
A1 & 2.51$\times10^{-5}$ & 0.04  & 5.56$\times10^{-7}$ \\
A2  &  9.99$\times10^{-13}$ &0.765   &  3.13$\times10^{-3}$ \\
A3 & 2.29$\times10^{-15}$ & 0.669  & 6.44$\times10^{-2}$ \\
\bottomrule
\end{tabular}
\caption{The results for the abundance and peak mass of PBHs, and the peak frequency of SIGWs for the models A1, A2 and A3.}
\label{tab:pbh}
\end{table}

\section{Scalar-induced gravitational waves}
Accompanied by the production of PBHs, the large primordial curvature perturbations at small scales can source the tensor perturbation at the second order after the horizon reentry to generate SIGWs.
In this section, we calculate the energy density of SIGWs generated in the models A1, A2, and A3.

\begin{figure}
\centering
\includegraphics[width=0.9\linewidth]{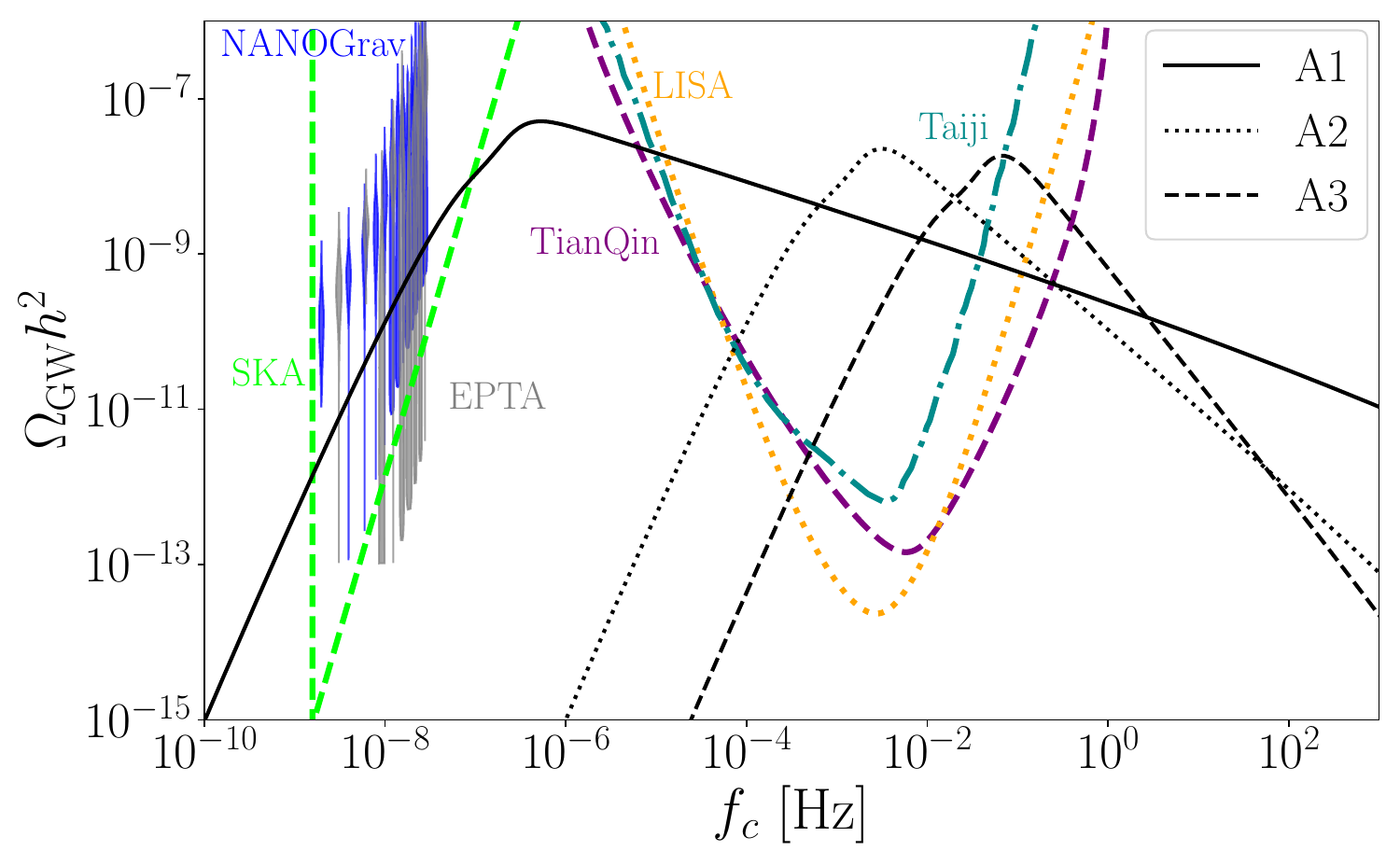}
\caption{The SIGWs generated from the models A1, A2, and A3. 
Black solid line, model A1;
black dotted line, model A2;
black dashed line, model A3. 
The blue and gray violins represent the NANOGrav 15-yr and EPTA DR2 data, 
respectively \cite{NANOGrav:2023ctt,EPTA:2023fyk}.
The green dashed curve denotes the SKA limit \cite{Moore:2014lga}.
The purple dashed curve denotes the TianQin limit \cite{TianQin:2015yph}, the orange dotted curve denotes the LISA limit \cite{LISA:2017pwj}, and the cyan dot-dashed curve denotes the Taiji limit \cite{Hu:2017mde}. 
The limits set by SKA, TianQin, LISA, and Taiji are the expected sensitivities when they are in operation in the future.}
\label{gws_img}
\end{figure}

The second-order tensor perturbation expressed in terms of the Fourier components is
\begin{equation}
h_{ij}(\bm{x},\eta)=\int\frac{d^3 \bm{k}}{(2\pi)^{3/2}}e^{i\bm{k}\cdot\bm{x}}[h_{\bm{k}}(\eta)e_{ij}(\bm{k})+\bar{h}_{\bm{k}}(\eta)\bar{e}_{ij}(\bm{k})],
\end{equation}
where $e_{ij}(\bm{k})$ and $\bar{e}_{ij}(\bm{k})$ are the plus and cross polarization tensors.
The power spectrum of the SIGWs is
\begin{equation}
\label{gwps}
\left\langle h_{\bm{k}}(\eta)h_{\bm{q}}(\eta)\right\rangle=\frac{2\pi^2}{k^3}\delta^{(3)}(\bm{k}+\bm{q})\mathcal{P}_h(k,\eta).
\end{equation}
The fractional energy density of SIGWs is \cite{Espinosa:2018eve}
\begin{equation}
\label{fomggw1}
\Omega_{\mathrm{GW}}(k)=\frac{1}{24}\left(\frac{k}{\mathcal{H}}\right)^{2}\overline{\mathcal{P}_h(k,\eta)}.
\end{equation}
Since GWs behave like radiation, the current fractional energy density of SIGWs is
\begin{equation}
\Omega_{\text{GW}0}(k)=\Omega_{\text{GW}}(k)\frac{\Omega_{r0}}{\Omega_{r}(\eta)},
\end{equation}
where $\Omega_{r}(\eta)$ is the fraction energy density of radiation and $\Omega_{r0}$ is its current value. 

Using the power spectra given in Fig. \ref{f+ps_img}, 
we calculate $\Omega_{\text{GW}0}$ for the models A1, A2 and A3 and the results are shown in Fig. \ref{gws_img}.
The value of the peak frequency $f_\text{c}$ where $\Omega_{\text{GW}0}$ takes the maximum value is shown in Table \ref{tab:pbh}. 
For the model A1,
even though the peak frequency of the SIGWs is around $10^{-7}$ Hz, 
the SIGWs have a broad spectrum and they can be detected in both the nano- and millihertz frequency bands.
The infrared parts of the SIGWs in the model A1 are consistent with the stochastic GW background detected by NANOGrav and EPTA.
For the models A2 and A3, the spectra of the SIGWs are relatively narrow.
The peak frequency of SIGWs for the model A2 is around $10^{-3}$ Hz
and the peak frequency of SIGWs for the model A3 is around $10^{-2}$ Hz. 
The SIGWs for the models A2 and A3 can be tested by future space-borne GW detectors such as
TianQin, Taiji, and LISA.

\section{Primordial Non-Gaussianity}
 
In this section, we calculate the non-Gaussianity of the primordial curvature perturbation $\hat{\zeta}$ by taking the model A1 as an example.
From the definition of the bispectrum $B_{\zeta}$ \cite{Byrnes:2010ft,Planck:2015zfm}
\begin{equation}
\label{bispec1}
\langle\hat{\zeta}_{k_{1}}\hat{\zeta}_{k_{2}}\hat{\zeta}_{k_{3}}\rangle=(2\pi)^{3}\delta^{(3)}(\bm{k}_{1}+\bm{k}_{2}+\bm{k}_{3})B_{\zeta}(k_{1},k_{2},k_{3}),
\end{equation}
we get the non-Gaussianity parameter $f_{\text{NL}}$ \cite{Byrnes:2010ft}
\begin{widetext}
\begin{equation}
\label{fnleq1}
f_{\text{NL}}(k_1,k_2,k_3)=\frac{5}{6}\frac{B_{\zeta}(k_1,k_2,k_3)}{P_{\zeta}(k_1)P_{\zeta}(k_2)+P_{\zeta}(k_1)P_{\zeta}(k_3)+P_{\zeta}(k_2)P_{\zeta}(k_3)},
\end{equation}
\end{widetext}
where $P_\zeta(k)=2\pi^2\mathcal{P}_\zeta(k)/k^3$, and the explicit form of the bispectrum $B_{\zeta}(k_{1},k_{2},k_{3})$ can be found in \cite{Hazra:2012yn,Zhang:2020uek}. In the squeezed limit $k_3\to 0$, there is a consistency relationship between the non-Gaussianity parameter $f_{\text{NL}}$ and the scalar spectral tilt $n_{s}$ \cite{Maldacena:2002vr}
\begin{equation}
\lim_{k_{3}\to 0}f_{\text{NL}}(k_1,k_2,k_3)=\frac{5}{12}(1-n_{s}).
\end{equation} 
Although the consistency relation was originally derived in the canonical single-field inflation with slow-roll conditions, 
it was then proved to be true for any inflation in which the inflaton is the only dynamical field during inflation \cite{Creminelli:2004yq}. 

\begin{figure}[htbp]
\centering
\includegraphics[width=0.9\linewidth]{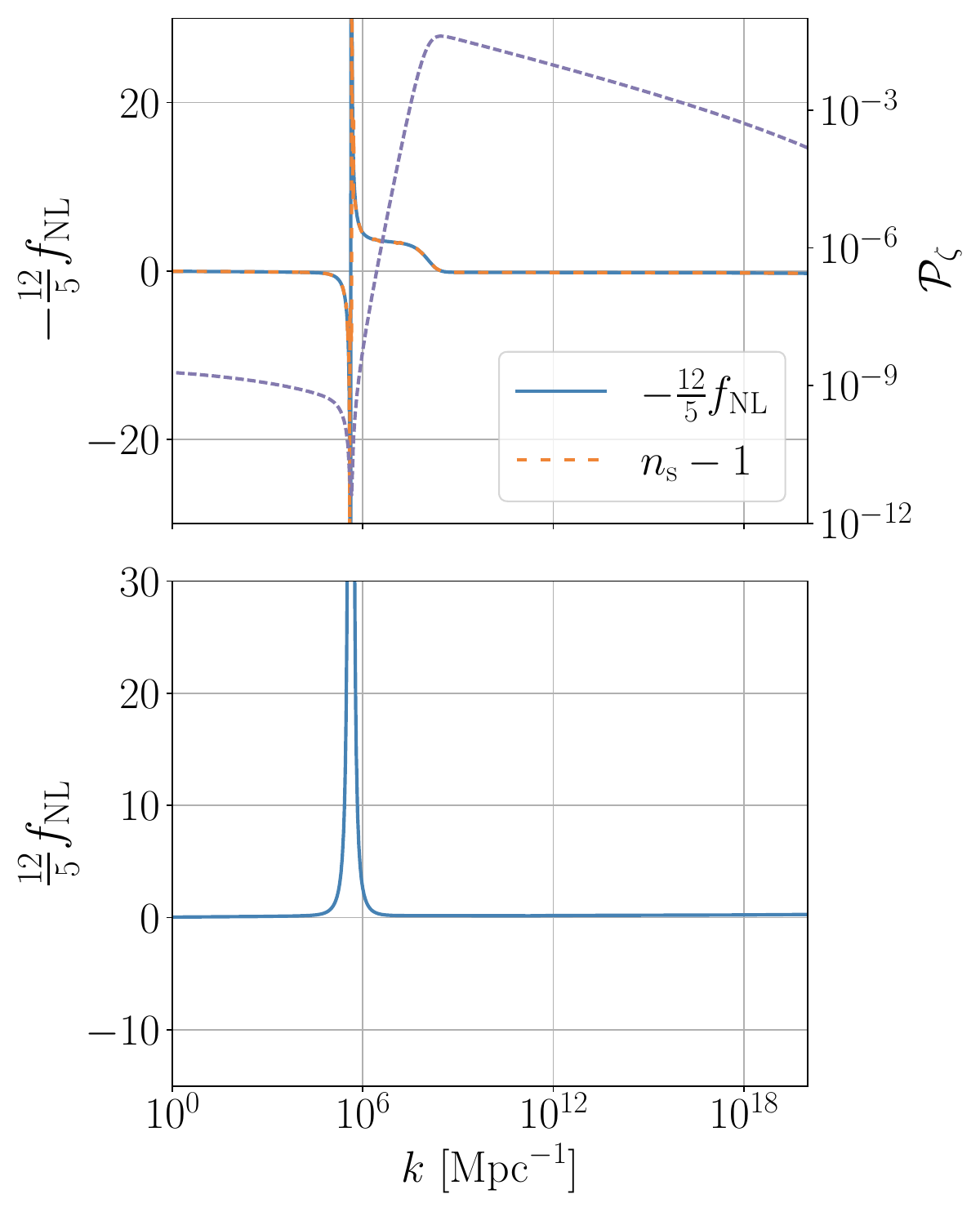}
\caption{The primordial scalar power spectrum $\mathcal{P}_{\zeta}$ and the non-Gaussianity parameter $f_{\text{NL}}$ for model A1. 
Upper: the purple dashed line represents the power spectrum $\mathcal{P}_{\zeta}$,  
the blue solid line represents $-\frac{12}{5}f_{\text{NL}}$ in the squeezed limit with $k_1=k_2=10^{6}k_{3}=k$, 
and the orange dashed line represents the scalar spectral tilt $n_{s}-1$. 
Lower: $\frac{12}{5}f_{\text{NL}}$ in the equilateral limit with $k_{1}=k_2=k_3=k$.}
\label{fnlA1_img}
\end{figure}

For the model A1, we numerically calculate $f_{\text{NL}}$ in the squeezed and the equilateral limits and 
the results are shown in Fig. \ref{fnlA1_img}. 
As shown in the upper panel of Fig. \ref{fnlA1_img}, the consistency relation holds in the squeezed limit.
In this model, the value of $f_{\text{NL}}$ is small when the power spectrum reaches the peak and is pretty large at scales around  $k\sim 10^{6}$ $\text{ Mpc}^{-1}$. 
The large values of $f_{\text{NL}}$ are due to the plunge of the power spectrum by several orders of magnitude. 
The change of the scalar spectral tilt near the peak is small, so the squeezed limit $f_{\text{NL}}$ is almost constant when the power spectrum climbs up and rolls down the peak.

Considering the non-Gaussianity of the primordial curvature perturbation $\zeta$, 
the PBH fractional energy density becomes \cite{Franciolini:2018vbk,Kehagias:2019eil,Atal:2018neu,Riccardi:2021rlf}
\begin{equation}
\beta=e^{\Delta_{3}}\beta^{G}.
\end{equation}
Since the mass of PBHs is almost monochromatic in the model A1, 
we can consider the correction on the peak scales only,
so the third cumulant $\Delta_3$ can be approximated as \cite{Zhang:2021vak}
\begin{equation}
\label{delta3eq1}
\Delta_{3}\approx \frac{23\,\delta^{3}_{c}}{\mathcal{P}_{\zeta}(k^{\text{peak}})}f_{\text{NL}}(k^{\text{peak}},k^{\text{peak}},k^{\text{peak}}).
\end{equation}
In the model A1, we have $\delta_c=0.4$, $\mathcal{P}_\zeta(k^\text{peak})=0.0338$,
and  $f_{\text{NL}}(k^{\text{peak}},k^{\text{peak}},k^{\text{peak}})=0.0754$,
so the third cumulant $\Delta_3=3.28>1$. 
Therefore the non-Gaussianity correction enhances the PBH abundance and this correction makes the production of PBHs easier.

In terms of the local-type non-Gaussianity parameter $f_\text{NL}$,
the primordial curvature perturbation can be parametrized as
\begin{equation}\label{fnl}
\zeta(\textbf{x})=\zeta^{G}(\textbf{x})+\frac{3}{5}f_{\text{NL}}(\zeta^{G}(\textbf{x})^{2}-\langle\zeta^{G}(\textbf{x})^{2}\rangle),
\end{equation}
where $\zeta^{G}(\textbf{x})$ is the Gaussian part of the curvature perturbation.
The power spectrum of the curvature perturbation is
\begin{equation}
\label{totpwr1}
\mathcal{P}_{\zeta}(k)=\mathcal{P}^{\text{G}}_{\zeta}(k)+\mathcal{P}^{\text{NG}}_{\zeta}(k),
\end{equation} 
where the non-Gaussianity correction to the power spectrum is \cite{Cai:2018dig,Zhang:2021vak}
\begin{equation}
\label{pwrnongaus1}
\mathcal{P}_{\zeta}^{\text{NG}}(k)=\bigg(\frac{3}{5}\bigg)^{2}\frac{k^{3}}{2\pi}f^{2}_{\text{NL}}\int d^3 \bm{p}\frac{\mathcal{P}_{\zeta}^{\text{G}}(p)}{p^{3}}\frac{\mathcal{P}^{\text{G}}_{\zeta}(|\bm{k}-\bm{p}|)}{|\bm{k}-\bm{p}|^{3}}.
\end{equation}
From Eq. \eqref{pwrnongaus1}, 
we expect that the non-Gaussianity correction to the power spectrum mainly comes from the value of $\mathcal{P}^\text{G}_\zeta$ around the vicinity of $k^\text{peak}$. 
Therefore, we use $f_\text{NL}$ in the equilateral limit at $k^\text{peak}$ as an estimator for the amplitude of the local-type non-Gaussianity parameter.
Using the results for the scalar power spectrum, 
we get $\mathcal{P}_{\zeta}^{\text{NG}}(k^\text{peak})=3.19\times 10^{-6}$.
Since the non-Gaussianity correction $\mathcal{P}_{\zeta}^{\text{NG}}(k)$ is negligible,
the effect of non-Gaussianity on the generation of SIGWs can be neglected.

\section{Conclusion}
Taking advantage of the existence of one inflection point in the third-order polynomial potential and inflationary attractors,
it was shown that the primordial scalar power spectrum in the models with the combination of the third-order polynomial and exponential $\alpha$ attractors 
can satisfy the CMB constraints at large scales and get enhancement at small scales to produce abundant PBHs and generate detectable SIGWs \cite{Frolovsky:2022qpg}.
For the exponential $\alpha$ attractors, 
the peak scale of the power spectrum $k^\text{peak}$ is about $10^{18} \text{ Mpc}^{-1}$,
and the PBHs produced from these models can reach asteroid size only.  
To produce PBHs with bigger mass, 
we combine the third-order polynomial and the polynomial attractor models to amplify the power spectrum at small scales. 
For the polynomial attractors,
$k^\text{peak}$ can be as small as $10^{14}\sim 10^{15} \text{ Mpc}^{-1}$,
and PBHs with the mass about $10^{17}$g can be produced. 
However, the mass of PBHs produced in these models is still too small. 
By adding a negative power-law term to the polynomials, 
abundant PBHs with different masses and the accompanying SIGWs with different peak frequencies are generated. 
For the three models considered in this paper, the peak scales $k^\text{peak}$ are $3\times 10^{8}$,
$1.7\times 10^{12}$, and $3.6 \times 10^{13} \text{ Mpc}^{-1}$, respectively;
the peak mass of PBHs are $2.51\times 10^{-5}M_\odot$, $9.99\times 10^{-13}M_\odot$,
and $2.29\times 10^{-15}M_\odot$, respectively;
the peak frequency of SIGWs are $5.56\times 10^{-7}$, $3.13\times 10^{-3}$, and $6.44\times10^{-2}$ Hz, respectively.
The PBHs with masses around $10^{-15}\textbf{--}10^{-12}$ $M_\odot$ can account for almost all DM.
The power spectrum of primordial curvature perturbations for the model A1 is broad and the power spectra for the models A2 and A3 are narrow,
so the SIGWs produced have both broad and narrow spectra. 
The infrared parts of the SIGWs in the model A1 can explain the stochastic GW background detected by NANOGrav and EPTA.
The non-Gaussianities of the primordial curvature perturbations in the squeezed and equilateral limits are calculated. 
We find that the non-Gaussianity correction enhances the PBH abundance
which makes the production of PBHs much easier,
but the effect of non-Gaussianity on the generation of SIGWs is negligible.

\begin{acknowledgments}
This research is supported in part by the National Key Research and Development Program of China under Grant No. 2020YFC2201504.
\end{acknowledgments}

%

\end{document}